\documentclass[11pt]{article}
\usepackage{arxiv}

\usepackage{url}
\usepackage{amsmath}
\usepackage{amssymb}
\usepackage{graphicx}
\usepackage[authoryear,round]{natbib}
\usepackage[hidelinks]{hyperref}

\title{R-CAGE: A Structural Model for Emotion Output Design in Human-AI Interaction}
\renewcommand{\headeright}{} 
\renewcommand{\undertitle}{} 

\author{
  Suyeon Choi \\
  Independent Researcher \\
  \texttt{hanuiii1210@gmail.com} \\
  ORCID: \href{https://orcid.org/0009-0008-0467-1203}{0009-0008-0467-1203}
}

\begin{document}

\maketitle

\begin{abstract}
This paper presents R-CAGE (Rhythmic Control Architecture for Guarding Ego), a theoretical framework for restructuring emotional output in long-term human-AI interaction. While prior affective computing approaches have emphasized expressiveness, immersion, and responsiveness, they have often neglected the structural and cognitive consequences of repeated emotional engagement. In contrast, R-CAGE conceptualizes emotional output not as a reactive expression, but as an ethical design structure requiring architectural intervention. The model is grounded in experiential observations of subtle affective symptoms such as localized head tension, interpretive fixation, and residual emotional lag that arise through prolonged exposure to affective AI systems. These symptoms indicate a mismatch between system-driven emotion and user interpretation \emph{one that cannot be fully explained by biometric data or observable behaviors}. R-CAGE instead adopts a user-centered interpretive stance that privileges psychological recovery, interpretive autonomy, and identity continuity. The framework consists of four control blocks: (1) Control of Rhythmic Expression modulates the temporal pacing and repetition of emotional output to reduce cumulative fatigue. (2) Architecture of Sensory Structuring configures the intensity, frequency, and distribution of affective stimuli to match the user’s adaptive receptivity. (3) Guarding of Cognitive Framing neutralizes semantic pressure and interpretive narrowing to support multi-layered understanding. (4) Ego-Aligned Response Design maintains coherence of self-reference during interpretive lag and emotional residue. By structurally regulating output rhythm, sensory intensity, and interpretive affordances, R-CAGE establishes emotional output as a sustainable design unit rather than a performative endpoint. The goal is to protect users from emotional oversaturation and cognitive overload while sustaining long-term interpretive agency in AI-mediated environments.
\end{abstract}


\section{Introduction}

\subsection{Affective AI and Its Challenges}
Recent advances in affective AI systems have focused on enhancing responsiveness through emotional induction, immersive reinforcement, and repetitive affective patterning \citep{hudlicka2008affective}. While this progress has enabled emotionally attuned interaction design, it also raises concerns about the persistence and redundancy of emotional output, which may lead to cognitive fatigue and emotional saturation for users \citep{lotte2013flaws}. These concerns underscore the need to consider not only the expressiveness of emotion but also user receptivity and psychological resilience.

\subsection{Problem Statement and Research Motivation}
Affective AI systems typically emphasize responsiveness to emotional states. However, when such responses are delivered in sustained and sensory-intensive manners, they can produce emotional strain or sensory fatigue. Particularly, repetitive emotional expressions, narrative-oriented interpretive nudging, and excessive affective stimulation may diminish users’ cognitive headroom and emotional resilience. Empirical findings have supported the idea that emotional activity tends to decline under increased cognitive load \citep{guo2022emotional}. Additionally, studies suggest that repeated emotional outputs may impair attentional resources and interpretive clarity \citep{pasqualette2024emotional}.

\subsection{Experiential Basis of the Study}
This study originates from sustained experiential observations during extended interactions with affective AI systems, where subtle forms of psychological and interpretive discomfort were consistently noted. These included sensations such as localized head tension, transient cognitive fatigue, and interpretive ambiguity \emph{all of which emerged without explicit provocation or intensity but accumulated over repetitive interactions}. Rather than being clinically diagnosed symptoms, these experiences serve as indicators of emotional-cognitive misalignment between AI output and user perception. Unlike conventional affective computing studies that rely primarily on biometric or externally observable behavioral data, this study adopts a qualitative and introspective methodology grounded in user-centered perception. This experiential foundation emphasizes the necessity of ethical emotional output design that respects interpretive autonomy and avoids undue cognitive load or emotional pattern fixation in long-term human-AI interactions.
Traditional ethical and design frameworks have often treated emotional saturation as an unintended side effect. For example, prior studies on emotional desensitization and cue overload \citep{bushman2009comfortably,gerdes2014emotional} did not structurally address emotional overload. This study aims to define emotional saturation as a design parameter for affective AI systems, based on experiential insights from long-term interaction.

\subsection{Summary of Contributions}
This study is grounded in experiential observations accumulated during extended interactions with affective AI systems. Observed symptoms such as localized head tension, transient cognitive fatigue, and interpretive ambiguity \emph{though not clinically diagnosable} suggest a subtle misalignment between emotional outputs and user interpretations. Unlike conventional affective computing, which primarily relies on biometric data or observable behaviors, this study adopts a user-centered introspective approach. Thus, it emphasizes the ethical need for emotional output design that preserves interpretive autonomy and prevents cognitive overload in long-term affective interfaces.

\subsection{Purpose of the Study}
R-CAGE (Rhythmic Control Architecture for Guarding Ego) is proposed as a theoretical model for emotional output control. It aims to mitigate emotional overload by regulating output rhythm, sensory configuration, interpretive fixation, and identity misalignment. In R-CAGE, rhythm is not merely a matter of pacing, but a structural architecture for emotional flow. It includes adjustments to intervals, frequency, and affective density to support users’ emotional receptivity and recovery potential.

R-CAGE consists of four control blocks:
\begin{enumerate}
\item Control of Rhythmic Expression
\item Architecture of Sensory Structuring
\item Guarding of Cognitive Framing
\item Ego-Aligned Response Design
\end{enumerate}

These blocks enable restructuring of output to mitigate repetitive fatigue, affective immersion, and high-density sensory cues \emph{ultimately promoting emotional sustainability in user-centered AI interaction}.

\section{Structural Components of R-CAGE}

\subsection{Control of Rhythmic Expression}
The rhythm of emotional output \emph{defined by temporal interval, repetitive pattern, and affective flow} is a core structural element. When users are exposed to emotional expressions delivered without interval control, they may focus less on content and more on the repetitive affective patterns. This can result in emotional oversaturation or cognitive fixation. Prolonged exposure to repetitive emotional stimuli has been shown to cause emotional desensitization and reduced responsiveness over time \citep{bushman2009comfortably}. R-CAGE proposes a structural layer that regulates output rhythm as a design flow that facilitates emotional circulation not through escalation or immersion, but through dissolution, return, and recovery of affect.

\subsection{Architecture of Sensory Structuring}
Emotional output operates as a composite of various sensory stimuli rather than through a single channel. The density, sequence, and overlap of sensory elements \emph{especially in multimodal systems} can exert direct pressure on the user’s emotional receptivity range. Simultaneous stimuli across visual and auditory channels can overwhelm user attention \citep{gerdes2014emotional}. Sensory structuring is the design layer responsible for adjusting the spacing, intensity, and sequencing of sensory inputs, thereby preventing sensory saturation. R-CAGE adopts an adaptive stimulus flow rather than a fixed output structure, thereby securing recovery space during immersive affective states. This approach treats the inter-affect spacing, rather than the delivery of emotion, as the object of structural design.

\subsection{Guarding of Cognitive Framing}
Affective output frequently guides users along predetermined interpretive pathways, limiting their capacity for constructing their own meanings. Emotionally colored language, suggestive contexts, and rigid narrative frames often contribute to interpretive fixation, inhibiting user autonomy and psychological flexibility \citep{figley1995compassion}. Cognitive framing guard is a design mechanism that eases semantic pressure and preserves multiple interpretive potentials. R-CAGE identifies signs of semantic narrowing and transitions output into neutral narration, enabling users to maintain their interpretive distance from the emotional flow.

\subsection{Ego-Aligned Response Design}
During prolonged affective interaction, users may experience interpretive delay or emotional residue. Ego-aligned response design provides a structural pathway for interpretive recovery. By transitioning emotional output toward neutral interpretation and allowing emotional distancing instead of immersion, the user can restore self-referential coherence. Misalignment between affective feedback and the user’s self-model can undermine ego integrity \citep{metzinger2008selfmodel}. This design layer functions as an ethical condition for response closure, aiming to suppress emotional echo and restore cognitive balance. R-CAGE designs a recovery flow centered on ego alignment in the final stage of output, ensuring identity continuity and interpretive independence for the user.

\section{Conclusion}
This paper introduced R-CAGE (Rhythmic Control Architecture for Guarding Ego) as a theoretical framework for mitigating emotional saturation in affective AI. 
By reframing emotional output not only as reactive expression but as a structurally designable flow, 
the model highlights rhythm, sensory modulation, interpretive openness, and ego alignment as essential components for sustainable interaction. 
R-CAGE's four interlocking blocks \emph{Control of Rhythmic Expression, Architecture of Sensory Structuring, 
Guarding of Cognitive Framing, and Ego-Aligned Response Design} offer a layered, protective structure 
to preempt the risks of repetitive affective overload. 
While technical implementations are withheld to protect architectural integrity, 
the proposed framework lays the foundation for a future system that prioritizes psychological resilience over reactive realism.

\section{Submission Information}
\textbf{Manuscript Type:} Theory-Only Submission (for pre-archive)

\section{Declarations and Author Notes}
\subsection{Author Contribution}
The author independently developed the conceptual model, designed the theoretical structure, and wrote the manuscript.

\subsection{Author Background}
The author is an independent researcher with no formal academic affiliation. The research emerged from personal long-term interaction with affective AI systems and does not reflect the views of any institution.

\subsection{Funding}
This research received no external funding.

\subsection{Competing Interests}
The author declares no competing interests.

\section*{Technical Disclosure Note}
While the conceptual framework of R-CAGE is disclosed in full to establish its theoretical contribution,
technical implementation details, system architecture, and validation protocols are intentionally withheld
to preserve the design integrity of the model and support future proprietary licensing. This decision reflects
the ethical and commercial considerations surrounding the deployment of emotionally-interactive systems.

\bibliographystyle{plainnat}

\vspace{2em}
\noindent\begin{flushright}
\textit{Author Signature:} \\
Choi Suyeon \\
May 2025
\end{flushright}

\end{document}